\documentclass[]{MathAppl22}
\usepackage[OT4,T1]{fontenc}
\usepackage[utf8]{inputenc}
\usepackage{polski}
\usepackage{graphicx}

\usepackage{amsmath}
\usepackage{amssymb}
\usepackage{filecontents}

\usepackage[english,polish]{babel} 
\usepackage{lastpage} 

\usepackage{etex} 

\usepackage{color} 
\usepackage{cite} 
\usepackage{algorithm}
\usepackage{algorithmicx}
\usepackage{algpseudocode}
\usepackage{listings}
\usepackage{bbm}

\RequirePackage[numbers]{natbib}

\usepackage[colorlinks=true]{hyperref}
  \hypersetup{
    pdftitle={Revisiting the Nowosiółka skull with RMaCzek}, 
    pdfauthor=Krzysztof Bartoszek, 
    colorlinks,
    urlcolor=blue,
    filecolor=magenta,
    citecolor=green, 
    linkbordercolor={1 1 1}, 
    citebordercolor={1 1 1},  
    urlbordercolor={ 1 1 1}  
  } 
 \RequirePackage[hyperpageref]{backref} 
    \renewcommand*{\backref}[1]{}  
    \renewcommand*{\backrefalt}[4]{
       \ifcase #1 
          No cited.
       \or
          Cited on p. #2.
       \else
          Cited on pp. #2.
       \fi}  
\newcommand{\orcid}[1]{\href{https://orcid.org/#1}{\includegraphics[scale=.05]{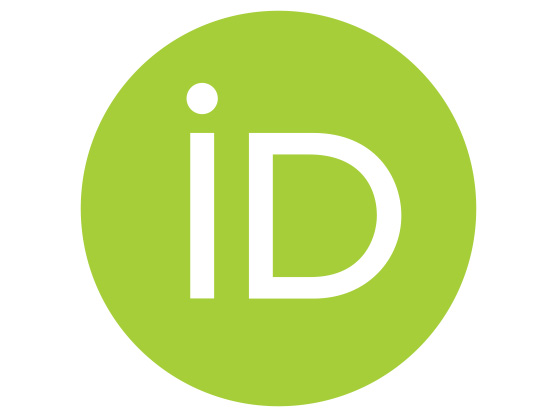}}}	
\newcommand{\orcidcode}[1]{\href{https://orcid.org/#1}{#1}}

\def\repo{http://wydawnictwa.ptm.org.pl/index.php/matematyka-stosowana/article/viewArticle}  

\pdfoutput=1
\usepackage{graphicx}
\usepackage{wrapfig}
\graphicspath{{./Figures/},{./Pictures/}}
\usepackage{multicol}
\firstpage{i}    
\LogoG{\includegraphics[width=0.18\textwidth]{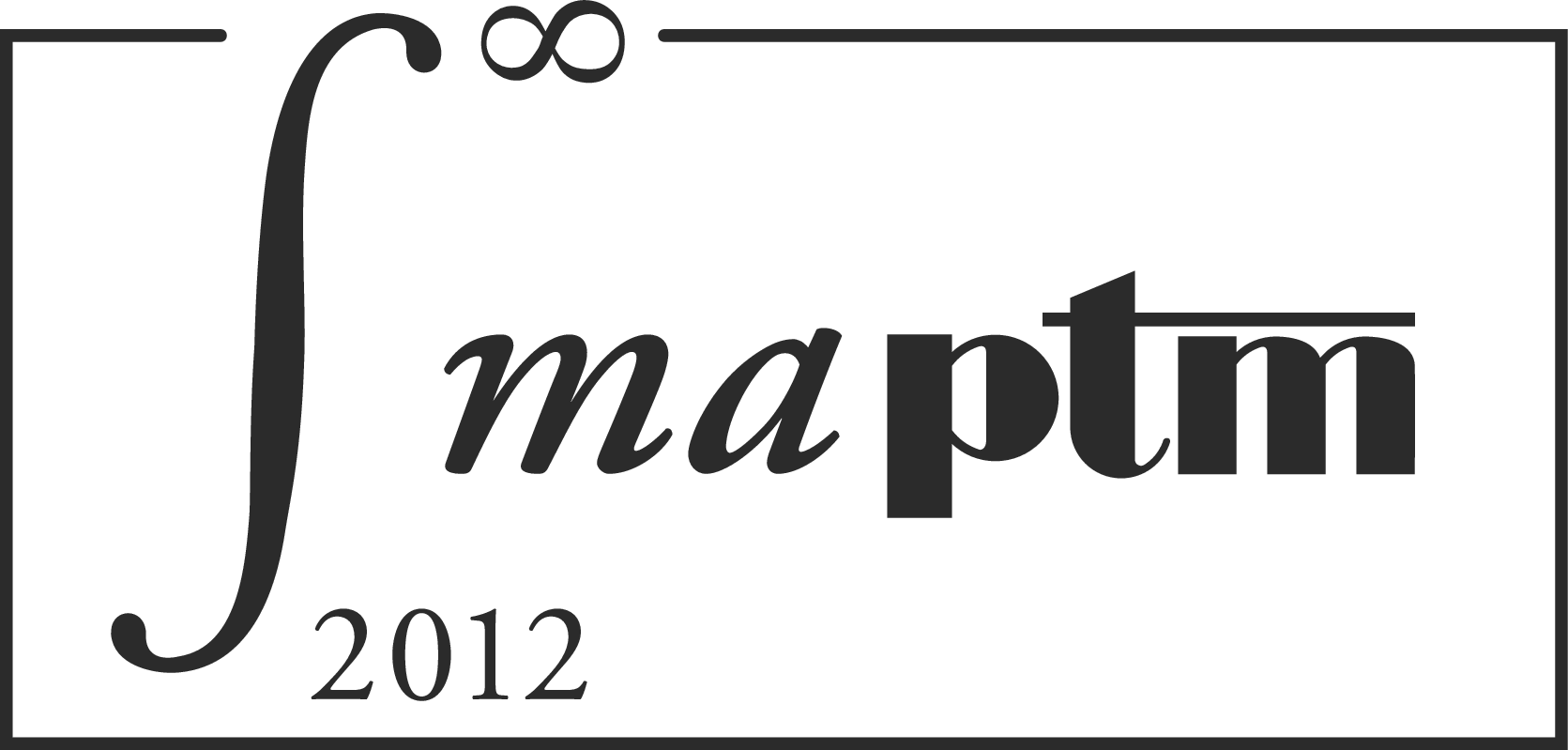}}
\volume{50}
\fasc{2}
\years{2022}
\LogoGMS{\includegraphics[width=0.18\textwidth]{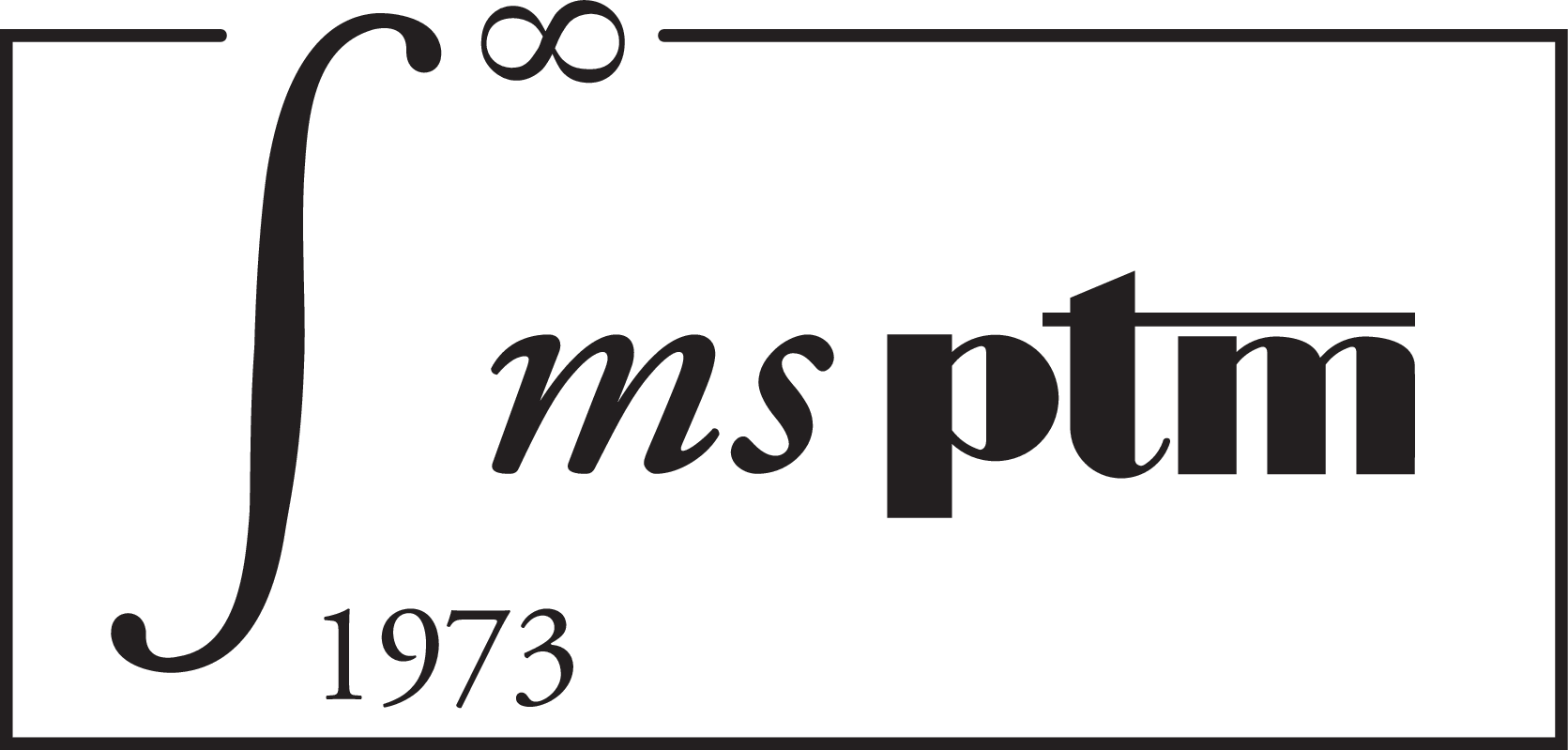}}
\volumeMS{35}
\numberMS{66}
\wwwtrue

\secnameMS{MATHEMATICS IN BIOLOGY \& MEDICINE}
\pages{255--266}
\received[28th of December 2022]{30th of September 2022}
\lastrevision{}
\logo{}{}{}
\def\doinum{10.14708/ma.v50i2.7164}

\title[Nowosi{\'o}{\l }ka skull and \pkg{RMaCzek}]{Revisiting the Nowosi{\'o}{\l }ka skull with \pkg{RMaCzek}}

\tpauthortrue 
\author[K. Bartoszek]{K.~Bartoszek\orcid{0000-0002-5816-4345}}
\thanks{\orcid{0000-0002-5816-4345}\orcidcode{0000-0002-5816-4345}\\
KB's research is supported by  the Swedish Research Council's (Vetenskapsr\aa det) grant no. $2017$--$04951$.
}
\affiliation{Link\"oping University}
\affiliation{Department of Computer and Information Science}
\address{The Division of Statistics and Machine Learning\\
\indent Link\"oping University, 581 83, Link\"oping, Sweden 
}
\city{Link\"oping}
\email{krzysztof.bartoszek@liu.se; krzbar@protonmail.ch}

\comm{Urszula Foryś} 
\newcommand{\be}{\begin{equation}} \newcommand{\ee}{\end{equation}}
\newcommand{\bd}{\begin{displaymath}} \newcommand{\ed}{\end{displaymath}}
\newcommand{\ba}{\begin{align}} \newcommand{\ea}{\end{align}}
\newcommand{\baa}{\begin{align*}} \newcommand{\eaa}{\end{align*}}
\newcommand{\ben}{\begin{enumerate}} \newcommand{\een}{\end{enumerate}}
\newcommand{\bi}{\begin{itemize}} \newcommand{\ei}{\end{itemize}}

\usepackage{algorithm}
\usepackage{algorithmicx}
\usepackage{algpseudocode}
\usepackage{listings}
\usepackage{bbm}

\newcommand{\pkg}[1]{\textbf{#1}}
\newcommand{\proglang}[1]{\textsf{#1}}
\newcommand{\code}[1]{\texttt{#1}}

\algnewcommand\algAnd{\textbf{and}~}
\algnewcommand\algOr{\textbf{or}~}


\begin{document}
\vspace{-5ex}
\setcounter{page}{255} 
\selectlanguage{english}\Polskifalse

\begin{abstract}
One of the first fully quantitative distance matrix visualization methods was proposed by Jan Czekanowski at the beginning of the previous century. Recently, a software package, \pkg{RMaCzek}, was made available that allows for producing such diagrams in \proglang{R}. Here we reanalyze the original data that Czekanowski used for introducing his method, and in the accompanying code show how the user can specify their own custom distance functions in the package.
\end{abstract}

\section{Introduction: Czekanowski's diagram}\label{secIntro}
Czekanowski's diagram is thought to be one of the first taxonomic and proximity visualization 
methods. It was proposed by the Polish anthropologist and statistician Jan Czekanowski in $1909$ \cite{JCze1909}.
In order to construct such a diagram one needs to be able to calculate the distance
between each pair of observations. Czekanowski used the average difference between the attributes
of two $d$--dimensional objects, $\vec{x}$ and $\vec{y}$,

\be\label{eqDD}
DD(\vec{x},\vec{y}) = \frac{1}{d}\sum_{r=1}^{d} \vert \vec{x}_{r}- \vec{y}_{r}\vert.
\ee
Then, one needs to solve a seriation problem \cite{MHahKHorCBuc2008}---find an arrangement of the observations
such that objects close together under the distance are close together when placed on a straight line.
Afterwords, one can represent the data as a matrix, where each cell is to represent the calculated distance 
between the row and
column object. Today heatmaps achieve this through a direct colour gradient, 
but in Czekanowski's times monochrome graphics were substantially
easier to produce on a large scale (but, e.g., in $1873$ a summary matrix of $40$ separate maps of Paris showing
various characteristics was presented in colour \cite{TLou1873,LWilMFri2009}). 
Hence, each cell is made up of a symbol representing the distance.
The number of possible symbols is limited---hence the distances are grouped.
The distance's range is divided into consecutive subintervals, with a unique
symbol assigned to each. For example, this can be a black dot with varying size---the smaller 
the distance the bigger the dot---see e.g. Fig. \ref{figJCze1909}.
Today, Czekanowski's method is classified under seriation, matrix reoordering and visualization methods. 
Closely, related to this are the matrix re--ordering and visualization methods proposed
by Bertin (e.g. \cite{JBer1967}). The main difference is that 
Czekanowski's approach works a similarity matrix, with rows and columns columns corresponding
to the same objects (with the same permutation of them),  while 
Bertin's focuses on presenting a matrix (e.g. containing the actual measurements for the different
objects) with possibly separate rearrangements of the rows and columns. 
In today's world, colour heatmaps and clustering methods are a common choice for visualizing matrices 
(be it for those containing similarity scores or actual measurements). This is also
due to their rapid development, alongside the increasing availability of 
of more and more sophisticated computer graphics. Currently a number of 
\proglang{R} \cite{R} heatmap packages are available on CRAN or
Bioconductor (e.g., \code{stats::heatmap()}, \pkg{ComplexHeatmap} \cite{ZGuREilMSch2016,ZGu2022}
\pkg{heatmap3} \cite{SZhaYGuoQSheYShy2014}, \pkg{heatmaps}, \pkg{pheatmap}, to name a few). 
For a thorough review, and history of this field we refer the reader to, e.g., \cite{ILii2010,LWilMFri2009}.

One can immediately notice that the creation of Czekanowski's diagram does not depend on the
actual way the distance was calculated---any function from the Cartesian product of
the observation's space with itself to the set of non--negative real numbers
(or even more generally to the space of symbols representing distances) would suffice.
Given Czekanowski's particular dataset (craniometric data of archaic humans)
the distance of Eq. \eqref{eqDD} sufficed. However, 
if one would have categorical observations, would want to take into account correlations
between attributes, weights of attributes or focus on alternative
aspects (as we will in Section \ref{secReanal}), then other distance functions could
be more appropriate. 

There are a number of previous software implementations of Czekanowski's diagram. The most well known
one was the \proglang{Visual Basic} \pkg{MaCzek} program \cite{ASolPJas1999}. 
Then, following encouragement from Miros{\l }aw Krzy{\'s}ko at the 
XXIV National Conference Applications of Mathematics in Biology and Medicine in Zakopane--Ko{\'s}cielisko, $2018$
and afterwords from Arkadiusz So{\l }tysiak, Albin V{\"a}sterlund \cite{AVas2019} implemented the 
\proglang{R} 
package \pkg{RMaCzek}.
The package allows for the creation of Czekanowski's diagram under user provided distance functions 
and seriation methods.
From our \textit{in--silico} experiments \cite{KBarAVas2020,AVas2019} it turned out that 
the most effective seriation method is based on finding a Hamiltonian path in the full graph of the observations
(with edge lengths equalling the distances between the observations) that is minimal \cite{ZBaretal2001}.

In this work we reanalyze the original data that introduced Czekanowski's diagram
to the world. We also provide scripts\footnote{\url{https://github.com/krzbar/RMaCzek_KKZMBM2020}
(with craniometric measurements)} 
that show how to use \pkg{RMaCzek}\footnote{\url{https://cran.r-project.org/web/packages/RMaCzek/}}. 
A key part of the scripts are the examples that show 
how to provide a custom user defined distance function
with arbitrary control parameters.
It is important to point out that comparing populations through cranial measurements
was dropped in the $1960$s thanks to the advancement in genetics
that showed that many morphological traits are environmentally controlled
and not genetically \cite{ASolTKoz2009}. We underline, that here we do not aim at
drawing conclusions concerning archaic human populations but at replicating 
analyses from over a century ago.

\section{The Nowosi{\'o}{\l }ka skull}
At the beginning of the previous century the debate on how humans developed to today
was fuelled by numerous fossil finds. Two hypotheses can be seen competing 
\cite{KSto1908}. The first one, ascribed to  Schwalbe \cite{KSto1908}, in modern language,
stated that \textit{Homo neanderthalensis} went completely extinct by the Paleolithic, not leaving
behind any intermediate forms to \textit{Homo sapiens}. The other, favoured
by Sto{\l }yhwo \cite{KSto1908} was that neanderthalic features survived the Paleolithic and could
also be present during the Era of History. He decided to disprove Schwalbe
through a semi--quantitative approach by comparing measurements of a 
Scythian, ca $30$--year old warrior's skull \cite{KSto1908} found in Nowosi{\'o}{\l }ka
(obviously \textit{H. sapiens}) with those of neanderthalic (\cite{KSto1908} after Schwalbe) skulls 
(those found in the caves of Neandertal, Spy and Krapina). 
Czekanowski \cite{JCze1909} performed a fully quantitative analysis 
of the craniometric data and found the Nowosi{\'o}{\l }ka 
skull to be placed in the \textit{H. sapiens}' cluster.

\section{Reanalysis with \pkg{RMaCzek}}\label{secReanal}
With the availability of the \pkg{RMaCzek} package we will attempt to replicate
both Sto{\l }yhwo's and Czekanowski's studies, do further data exploration and
see whether any of two hypotheses concerning the survival of neanderthalic features is better supported. 
Sto{\l }yhwo presented measurements of $47$ cranial features, some of them multivariate,
from $31$ human skulls. While this might seem a lot, most of the measurements
are missing, and most skulls have only a few features measured. For some skulls a single
value is provided for a given feature, while for others the range of the feature is provided.
Finally, in some cases it is not possible to assign a feature to any particular skull,
it is just written ``piece of skull from Krapina''. For the Nowosi{\'o}{\l }ka skull
all features are measured. Then, the Nowosi{\'o}{\l }ka skull's measurements
are compared with neanderthalic (called \textit{Homo primigenius} in \cite{KSto1908}, today this
nomenclature is deprecated and \textit{H. neanderthalensis} is used instead) ones. Each
feature is classified either as not--different, similar to or different from \textit{H. primigenius}'s
respective feature. Unfortunately it is not stated explicitly what computational procedure
is employed for this classification. 
Czekanowski, employed the approach described in Section \ref{secIntro},
with the distance of Eq. \eqref{eqDD}, to a subset of $13$ skulls, using $27$ of the features. 
He cites \cite{KSto1908} as the data source. The choice of skulls and features is not 
explained in \cite{JCze1909}, but it seems that the skulls with more measurements were chosen
and the features are such that they are present in the Neandertal, Br{\"u}x (\textit{H. sapiens} representative)
and Nowosi{\'o}{\l }ka skulls. Czekanowski does not describe the rationale for his arrangement
of the skulls but it could be guided by the timeline presented in \cite{GSch1906}, on p. $14$.

We first digitalized all the data presented in \cite{KSto1908} 
and they are made available 
alongside the \proglang{R} scripts for this work.
For further analyses we keep to the $13$ skulls of \cite{JCze1909}.
In these $13$ skulls $64.6\%$ of measures variables considered by Sto{\l }yhwo are missing,
if we restrict ourselves to the $27$ variables considered in \cite{JCze1909}
we have ``only'' $29.9\%$ missingness. However there are still substantial missing value
levels amongst the individual skulls:
Spy I $(7.4\%)$, Spy II $(14.8\%)$, Krapina C  $(40.7\%)$, Krapina D  $(59.3\%)$,
Neandertal  $(0\%)$, Gibraltar $(48.1\%)$, Pithecanthropus $(22.2\%)$, Kannstatt $(18.6\%)$,
Galey Hill $(55.6\%)$, Brunn $(55.6\%)$, Br\"ux $(0\%)$, Egisheim $(66.7\%)$ 
and Nowosi{\'o}{\l }ka $(0\%)$.
It was noticed \cite{ASolPJas1999} that observations that have missing values
on more than half the variables should not be used, as they may be close
to multiple, often very different from each other observations. Hence in our 
data set missingness is a potential serious problem.

\begin{figure}[!h]
  \begin{center}
\includegraphics[width=0.495\textwidth]{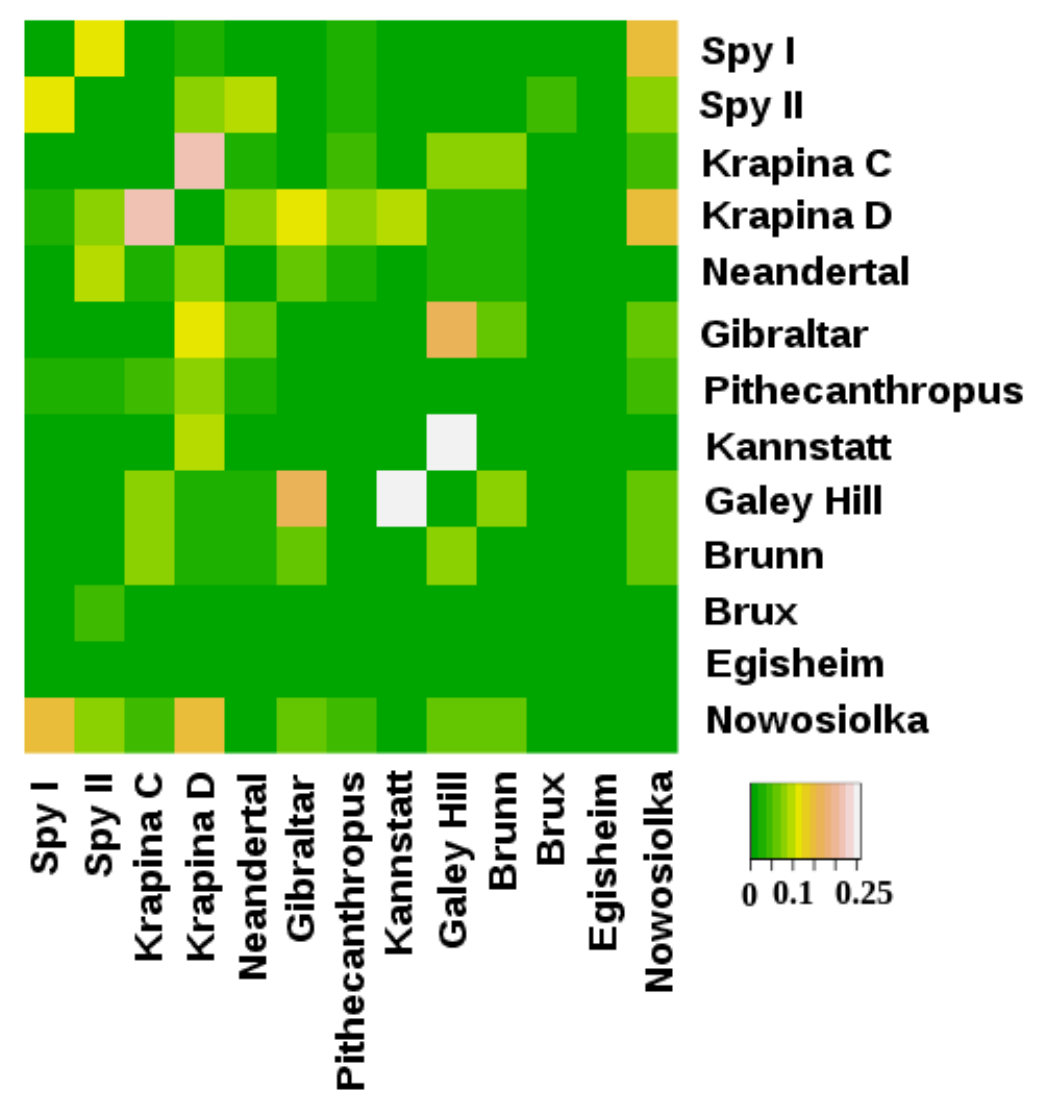}  
\includegraphics[width=0.495\textwidth]{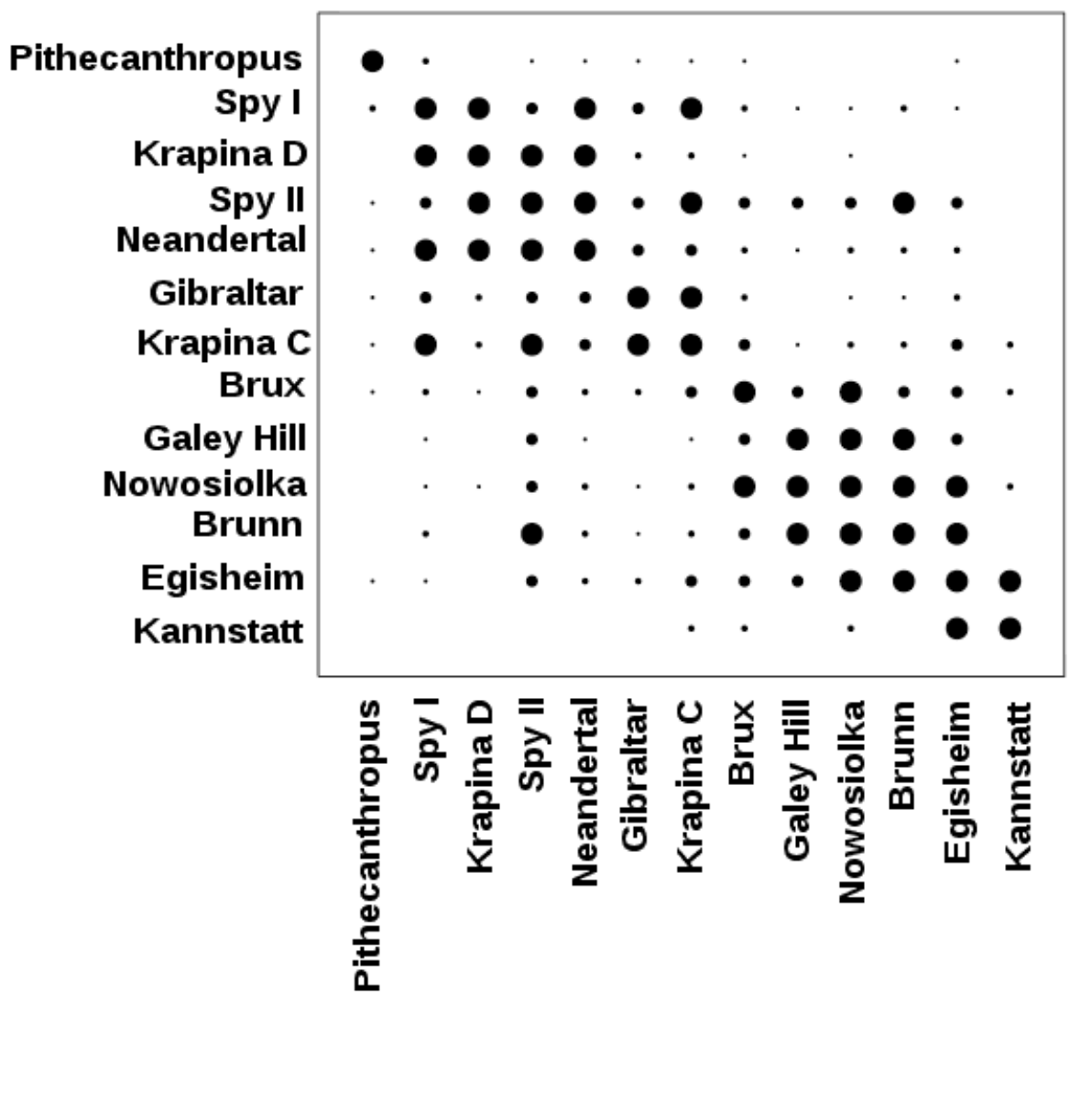}  
  \end{center}
    \vspace{-20pt}
  \caption{Left: absolute relative difference between skulls' distance
matrix found in \cite{JCze1909} and the distance matrix calculated 
using \pkg{RMaCzek}, right: Czekanowski's diagram found using
\pkg{RMaCzek}.}
   \label{figJCze1909}
\end{figure}

Then, our first goal was to replicate 
the distance matrix (Tabelle II in \cite{JCze1909}) and diagram (Tabelle III in \cite{JCze1909}) between in the skulls
that can be found in \cite{JCze1909}. We illustrate this in Fig. \ref{figJCze1909}.
We can see that our derived distance matrix differs a bit from the one Czekanowski
presented in \cite{JCze1909}. However, all but a few cells differ by less than $4\%$. 
Certainly a substantial amount of the difference
can be attributed to rounding errors, Czekanowski had to commit them when
calculating manually or with a mechanical calculator, and also his distances are presented up to the third
decimal point. However, it is difficult to discover what the larger 
discrepancies are due to. We do not know if he only used the
measurements presented in \cite{KSto1908}, or maybe some additional ones.
Furthermore, we also do not know how variables with only an upper and lower bound 
provided were treated.
Here we take the arithmetic
average of the maximum and minimum of the range.
It is worth pointing out that in Czekanowski's
original distance matrix there is an obvious typo in the
distance between the Neandertal and Galley Hill skulls. In Fig. 
\ref{figJCze1909} we plot the resulting Czekanowski's diagram
for the best found arrangement 
of the skulls by \pkg{RMaCzek}.
In \cite{KBarAVas2020} we provided the between--skulls distance
matrix from \cite{JCze1909} to \pkg{RMaCzek} and obtained
a better (under all objective functions) permutation 
 than in 
\cite{JCze1909}, however all the qualitative conclusions were the
same. 
Here, we find the same arrangement, 
derived directly from the raw measurements
and we obtain the same figure as in \cite{KBarAVas2020}. 

We now turn to finding the placement of the Nowosi{\'o}{\l }ka skull with respect
to the other skulls. We consider three distances---the DD distance of Eq. \eqref{eqDD},
the squared Euclidean distance and counting the number of variables 
supporting different classification (\textit{H. neanderthalensis} or 
\textit{H. sapiens}). The first and the last one are those considered by
Czekanowski and Sto{\l }yhwo, respectively, and hence relevant for our attempt
to replicate the results. We then take the Euclidean distance function
for comparison. 
We also try out a number 
of other options. We keep angles measured in degrees and also convert them to radians.
Some of the variables are ratios of other variables---they are dependent
and this could interfere with the seriation procedure. 
Hence, apart from all the
variables, we also perform analyses with the ratios removed and alternatively
where the ratio variables are kept, but their components are removed.
We also take into consideration
all the variables, or only the $27$ considered in \cite{JCze1909}. 
We take all observations and also removed those with more than $50\%$ missingness.
We also normalize, mean centre and divide by standard deviation, all the variables.

Furthermore, we wanted to see if we can replicate the conclusion of \cite{KSto1908}, that 
Nowosi{\'o}{\l }ka is more similar to neanderthalic skulls. To do this we needed to
implement a distance function that could mimic the table on p. $25$ in \cite{KSto1908}.
Sto{\l }yhwo considered whether each variable measured in the Nowosi{\'o}{\l }ka skull 
was the same, similar or different from \textit{H. primigenius} ones. To mimic
this we employ the following procedure. We take two focal skulls---Neandertal 
(representing the \textit{H. neanderthalensis} clade) 
and Br{\"u}x (representing the \textit{H. sapiens} clade).
We take this pair as they have measurements on all $27$ variables considered
in \cite{JCze1909}. Then, to calculate the distance between
observations \code{x} and \code{y}, with \texttt{v1} and \texttt{v2}
being the measurements of the  representatives of the two clades, 
the below \proglang{R} procedure is used.

{\small
\begin{lstlisting}
x1<-abs(x-v1);x2<-abs(x-v2);y1<-abs(y-v1);y2<-abs(y-v2)
v_x<-x1-x2;v_y<-y1-y2
qx<-quantile(v_x,probs=c(1/3,2/3),na.rm=TRUE)
qy<-quantile(v_x,probs=c(1/3,2/3),na.rm=TRUE)
xd<-rep(NA,length(v_x));yd<-rep(NA,length(v_y))
xd[which(v_x<=qx[1])]<-0;xd[which(v_x>qx[2])]<-2
xd[intersect(which(v_x>qx[1]),which(v_x<=qx[2]))]<-1
yd[which(v_y<=qy[1])]<-0;yd[which(v_y>qy[2])]<-2
yd[intersect(which(v_y>qy[1]),which(v_y<=qy[2]))]<-1
distxy<-mean(abs(xd-yd),na.rm=TRUE)

\end{lstlisting}
}
\noindent
We may recognize that what is done is that for both
\code{x} and \code{y} each variable is classified as
being closer to the respective variable in 
\code{v1} or  \code{v2} (or in--between).
Then, we take the average of how these patterns differs between \code{x} and \code{y}.
Importantly, under this distance normalization of the variables is not done.

\begin{figure}
  \begin{center}
\includegraphics[width=0.495\textwidth]{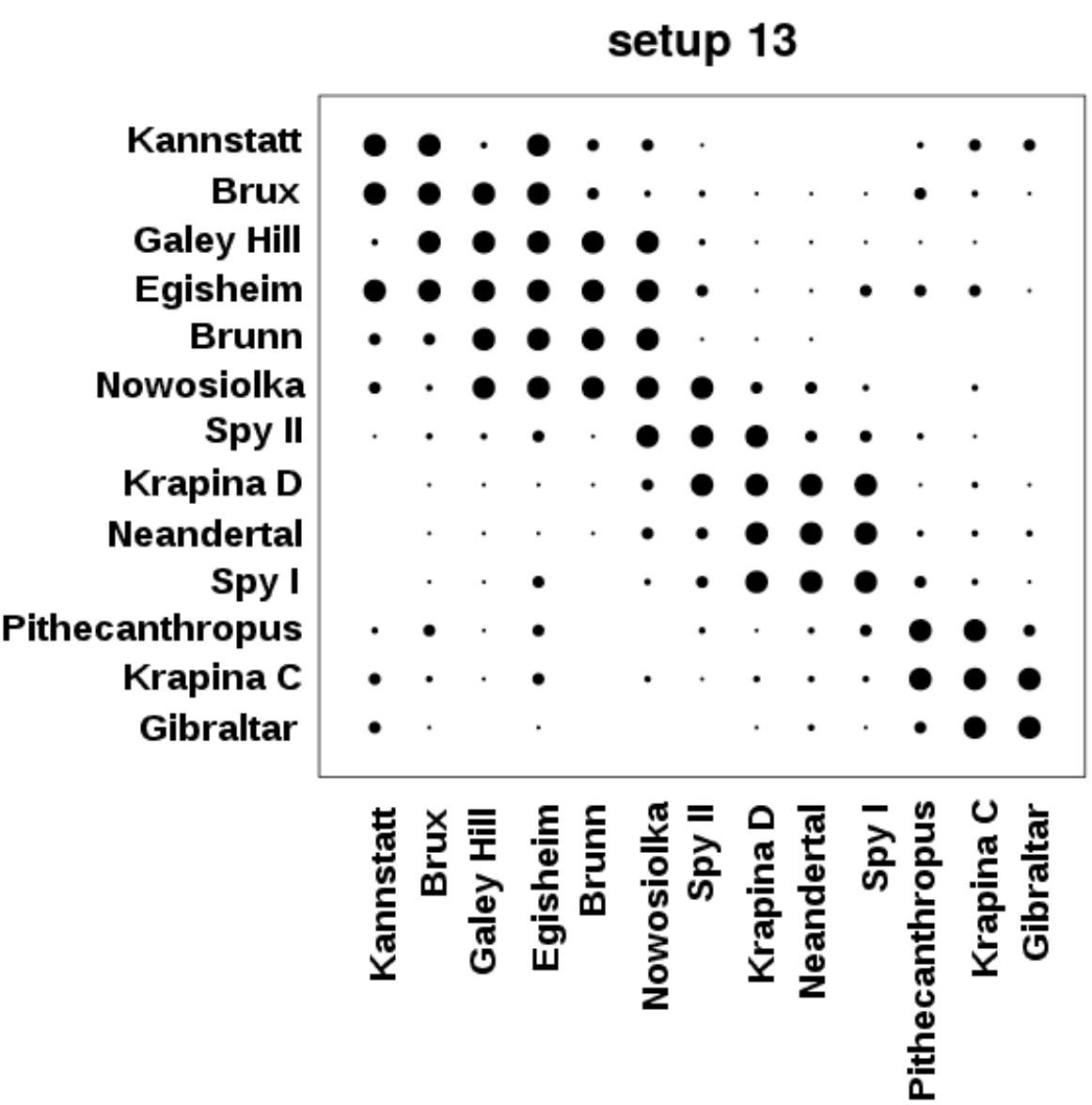}
\includegraphics[width=0.495\textwidth]{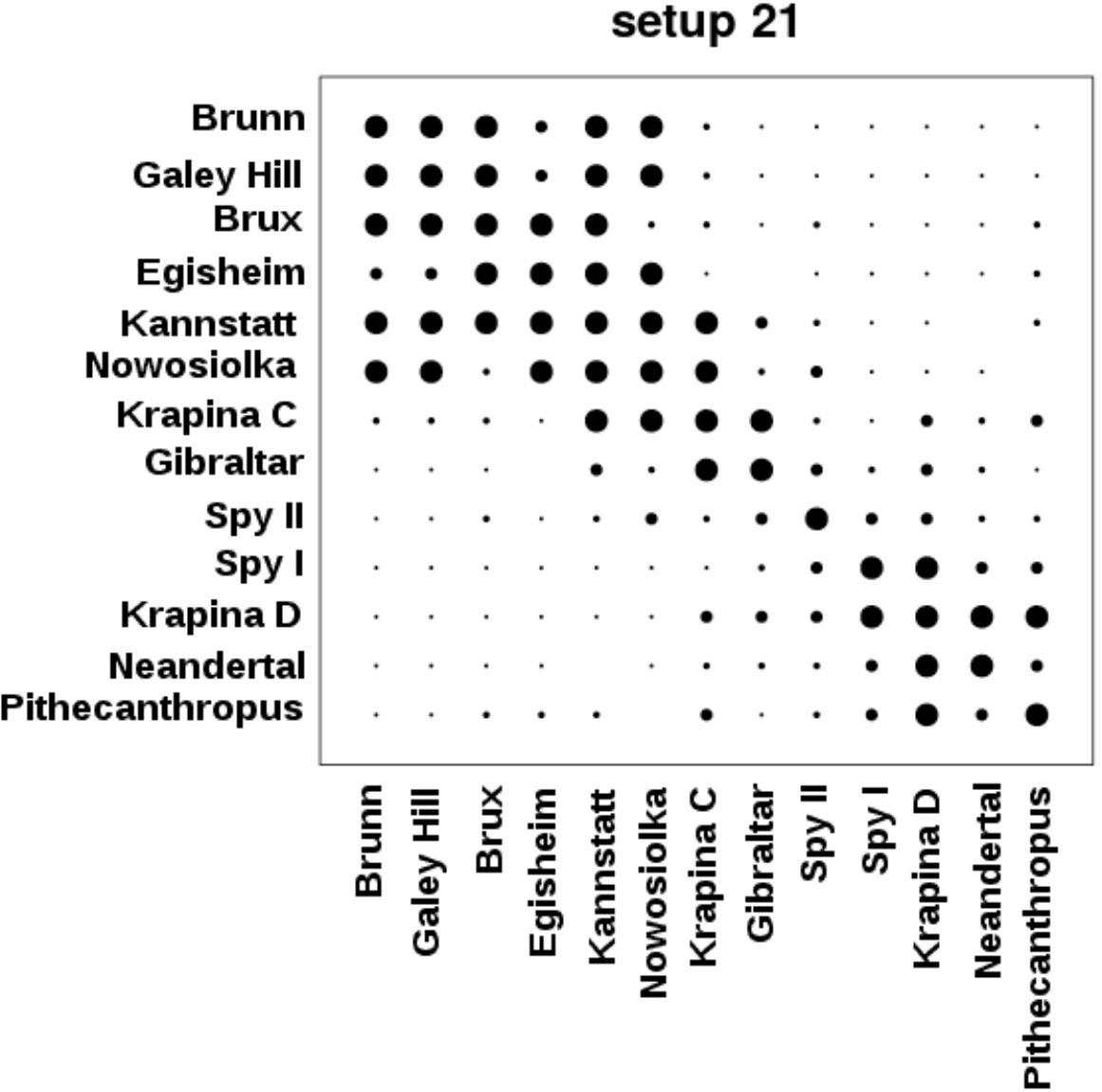}\\
\includegraphics[width=0.495\textwidth]{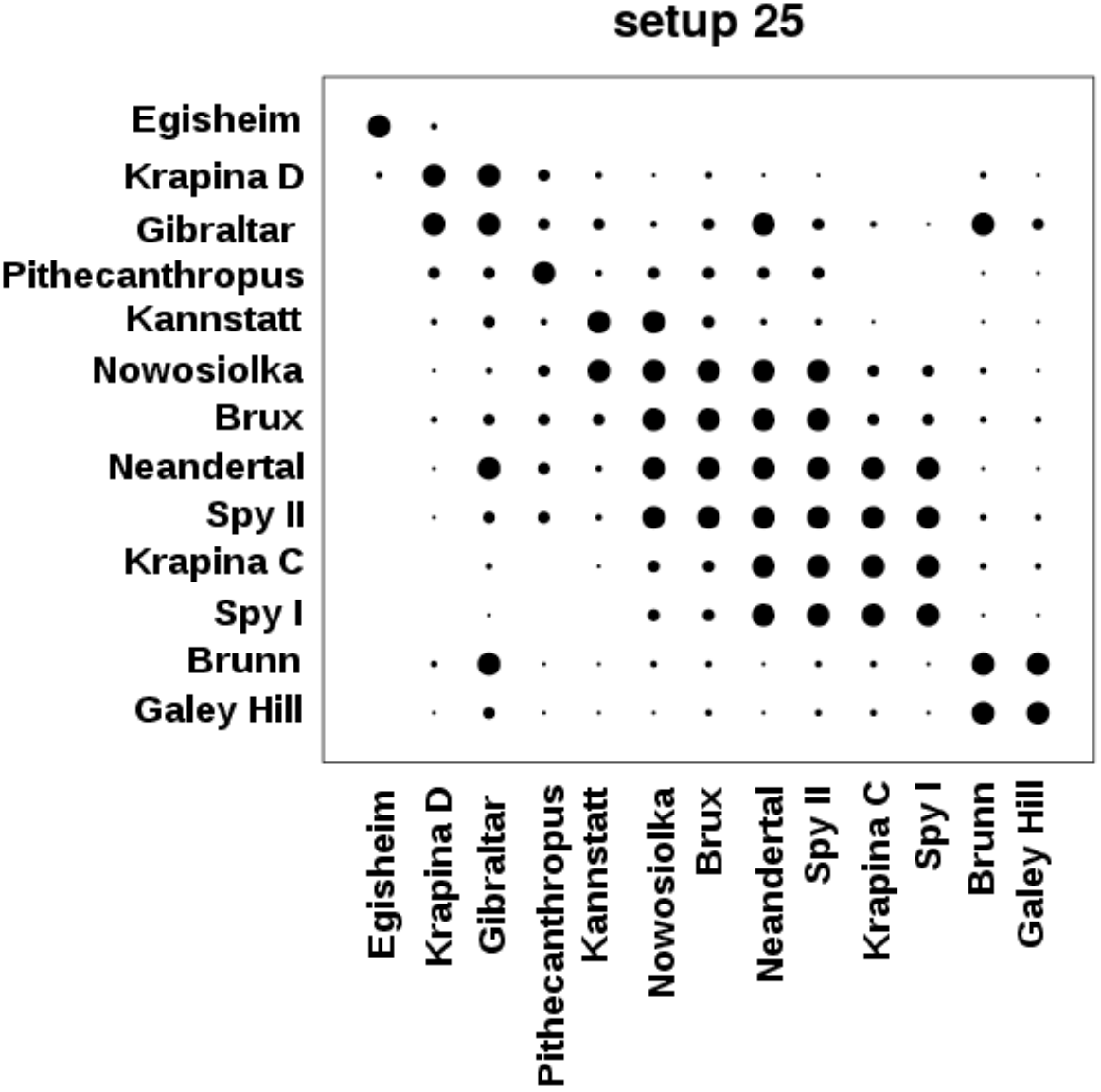} 
\includegraphics[width=0.495\textwidth]{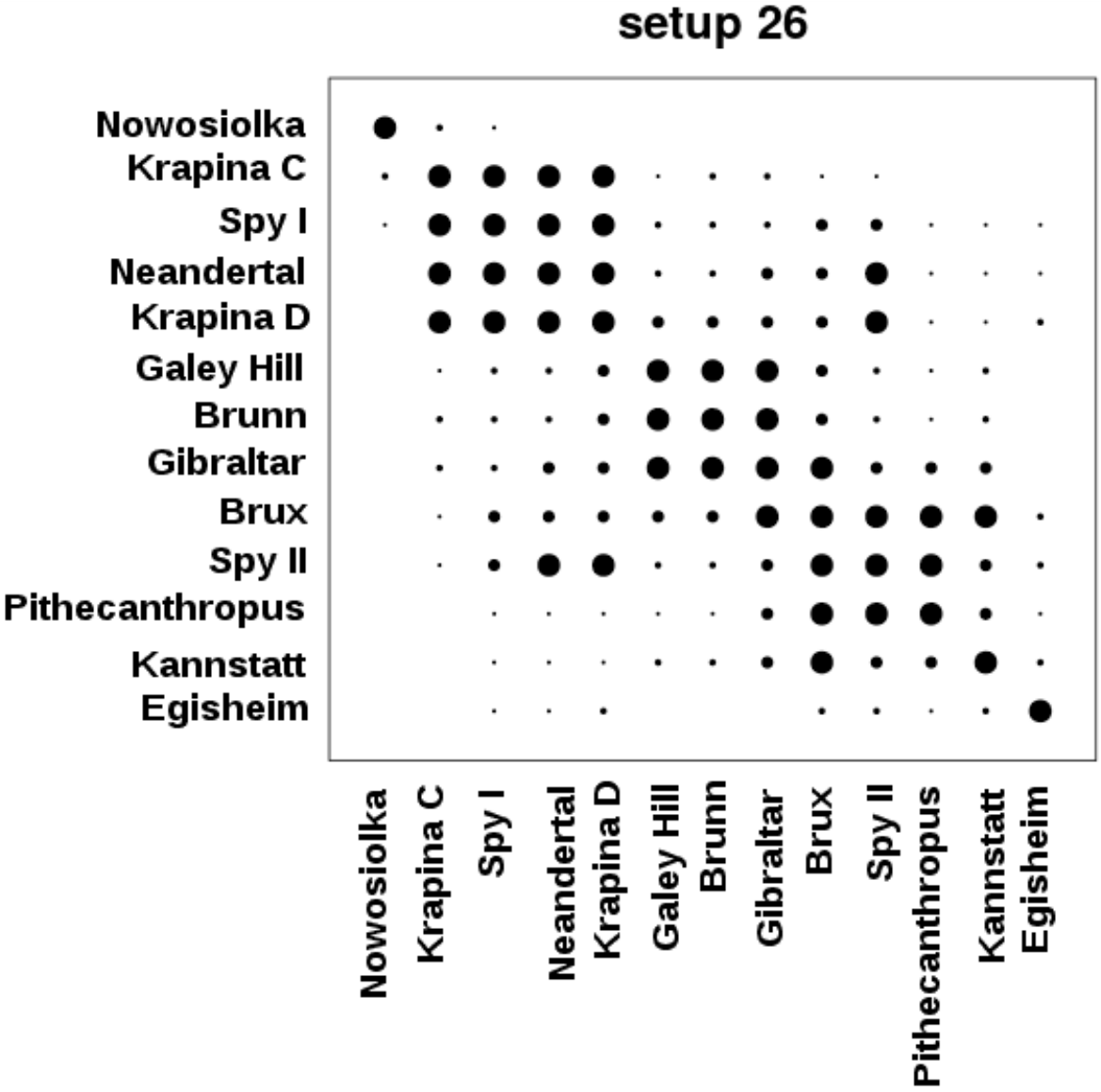} \\
 \end{center}
 \vspace{-20pt}
  \caption{Exemplary (part 1) Czekanowski's diagrams, where the Nowosi{\'o}{\l }ka
skull was placed closer to the  \textit{H. neanderthalensis} skulls
than in the original analysis by Czekanowski \cite{JCze1909} and \pkg{RMaCzek} 
with Czekanowski's settings.
}\label{figNowsiolkaCloser_1}
\end{figure}

\begin{figure}
  \begin{center}
\includegraphics[width=0.495\textwidth]{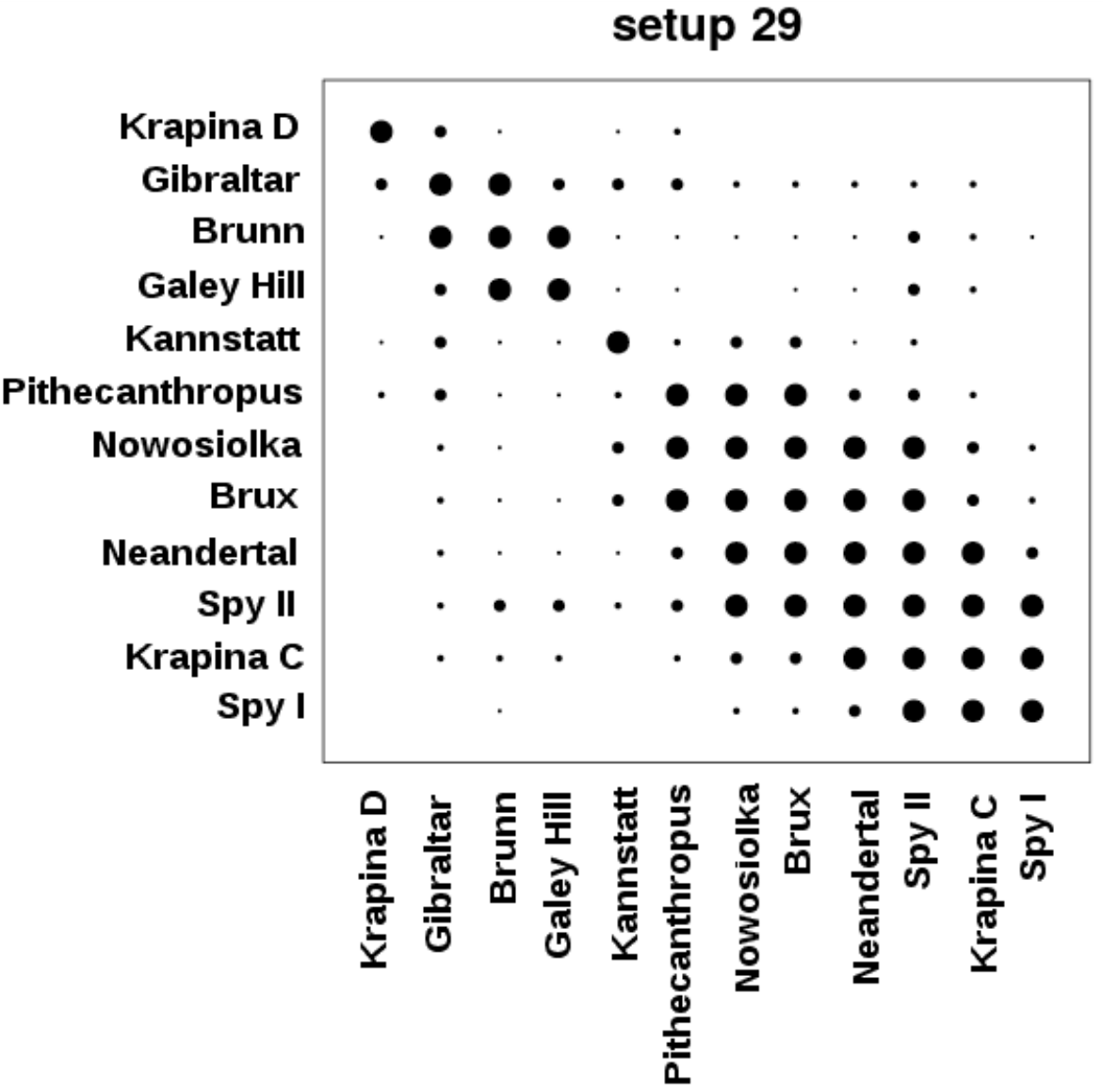}
\includegraphics[width=0.495\textwidth]{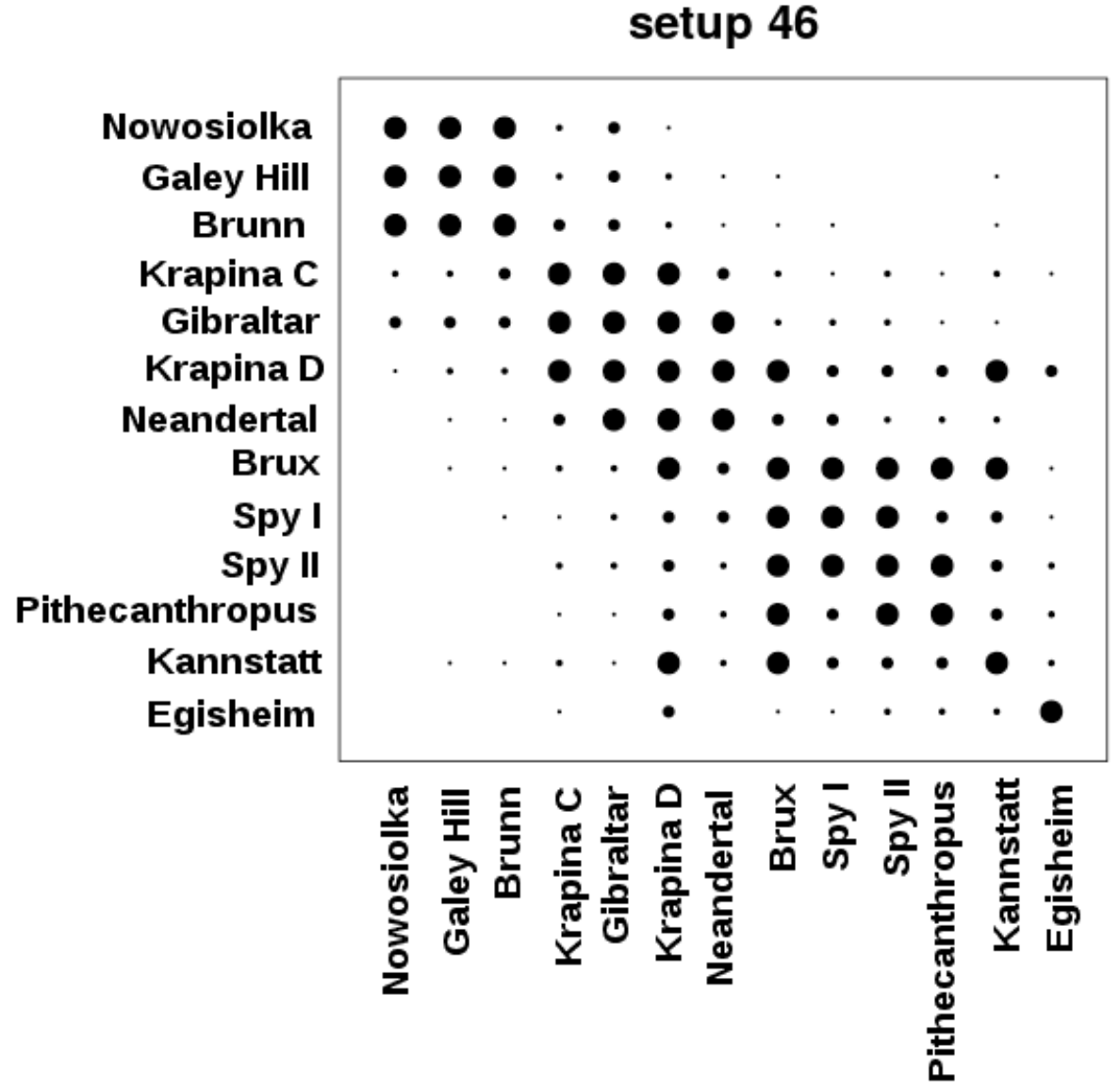} \\
\includegraphics[width=0.495\textwidth]{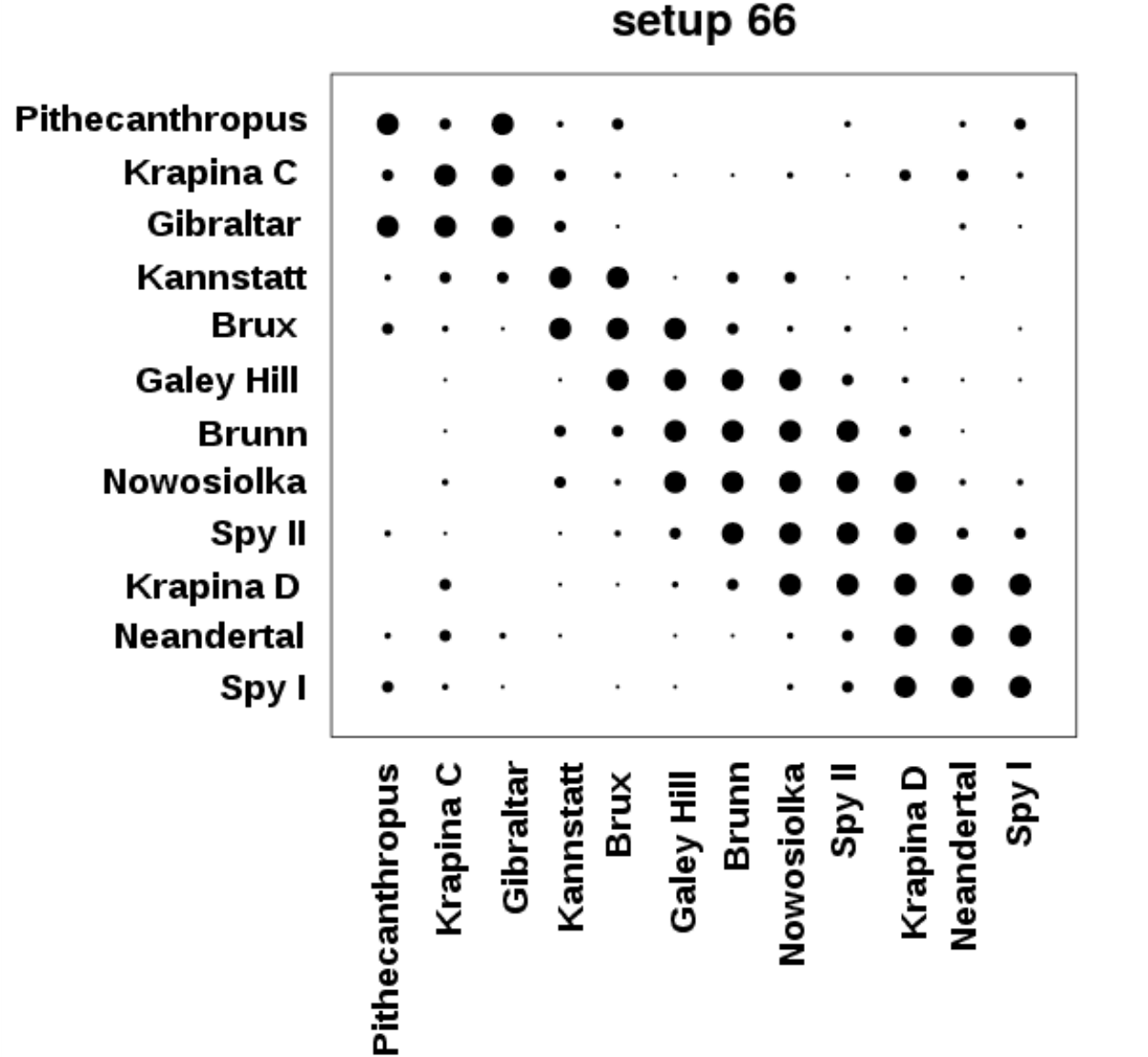}
\includegraphics[width=0.495\textwidth]{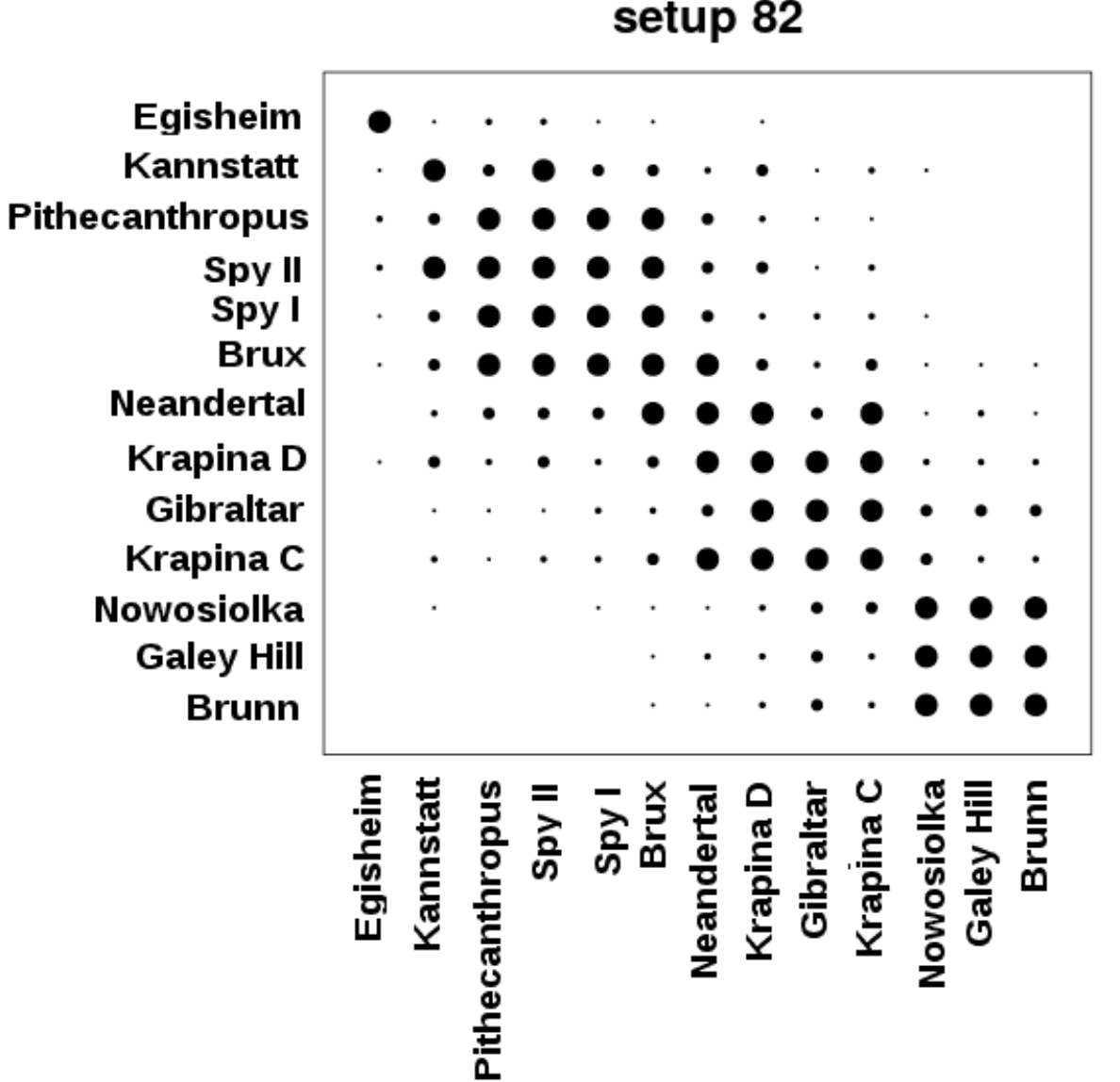} \\
\includegraphics[width=0.495\textwidth]{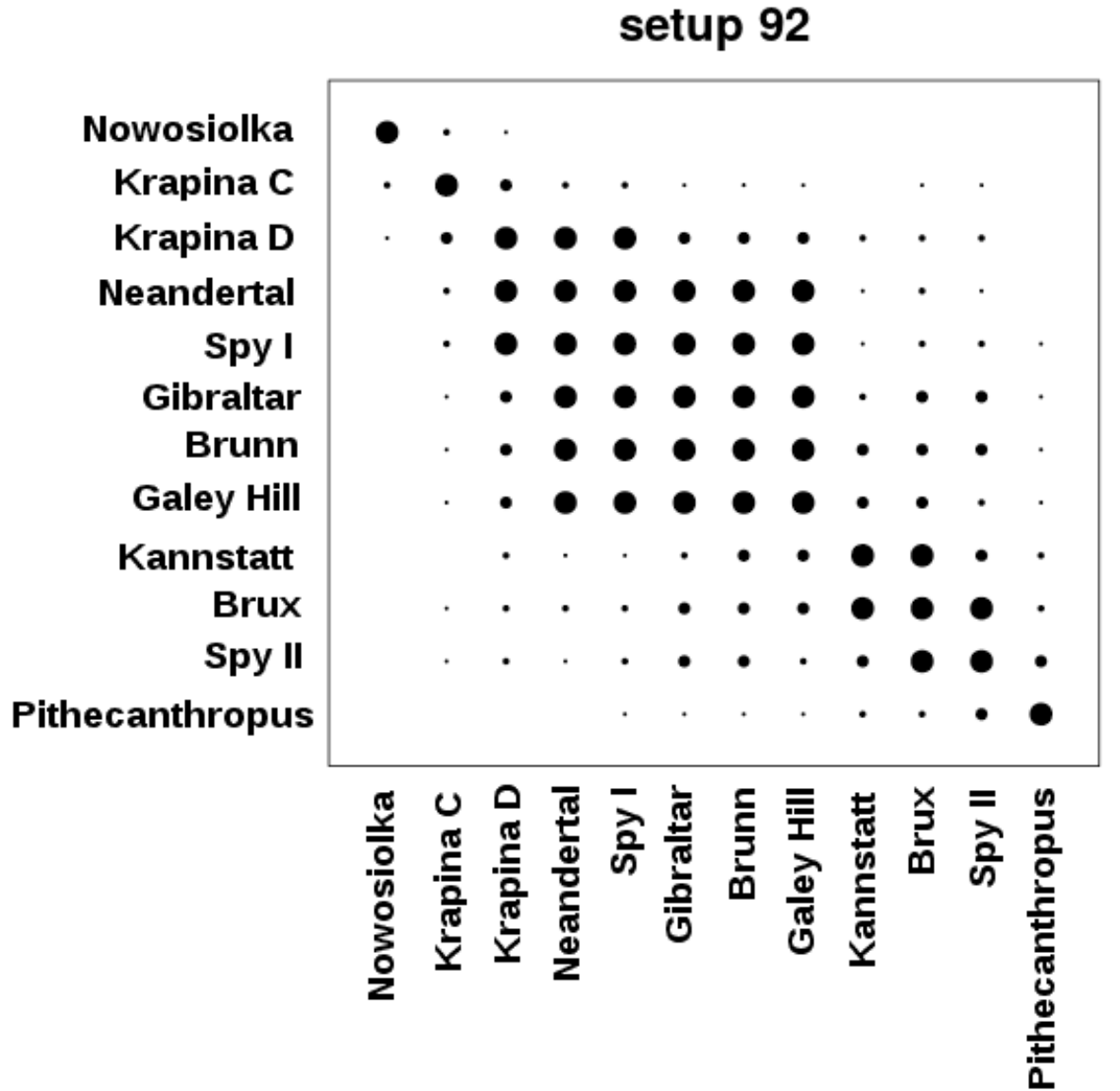}
 \end{center}
 \vspace{-20pt}
  \caption{Exemplary (part 2) Czekanowski's diagrams, where the Nowosi{\'o}{\l }ka
skull was placed closer to the  \textit{H. neanderthalensis} skulls
than in the original analysis by  Czekanowski \cite{JCze1909} and \pkg{RMaCzek} 
with Czekanowski's settings.
}\label{figNowsiolkaCloser_2}
\end{figure}

All together this resulted in $96$ possible setups (with resulting arrangements of the 
skulls). We went through this list manually 
and in $65$ setups the Nowosi{\'o}{\l }ka skull was placed closer to the \textit{H. neanderthalensis} skulls
than in the original result presented in Fig. \ref{figJCze1909}.
These cases can be divided as those
where the Nowosi{\'o}{\l }ka skull was: 
placed on the boundary between \textit{H. sapiens} and \textit{H. neanderthalensis} skulls
($11$ setups, exemplary setup number $13$ in Fig. \ref{figNowsiolkaCloser_1});
separated from the \textit{H. sapiens} skulls by
the Kannstatt skull (which in the original presentation can be thought to be a singleton) 
and followed by \textit{H. neanderthalensis} skulls
($6$ setups, exemplary setup number $21$ in Fig. \ref{figNowsiolkaCloser_1});
on the border with \textit{H. neanderthalensis} skulls 
of a partial \textit{H. sapiens} skulls set 
($2$ setups, exemplary setup number $82$ in Fig. \ref{figNowsiolkaCloser_2}); 
on the border of the whole \textit{H. sapiens} skulls set which was placed inside the 
\textit{H. neanderthalensis} skulls 
($4$ setups, exemplary setup number $66$ in Fig. \ref{figNowsiolkaCloser_2});
jointly with the Kannstatt and Br{\"u}x skulls placed amongst \textit{H. neanderthalensis} skulls
($23$ setups, exemplary setup number $25$ in Fig. \ref{figNowsiolkaCloser_1});
jointly with the Br{\"u}x skull placed amongst \textit{H. neanderthalensis} skulls
($1$ setup, number $29$ in Fig. \ref{figNowsiolkaCloser_2});
a singleton followed by \textit{H. neanderthalensis} skulls
($18$ setups, exemplary setups number $26$ and $92$ in Figs. \ref{figNowsiolkaCloser_1} and \ref{figNowsiolkaCloser_2}).
Unrelated to this, but interestingly, the Br{\"u}x skull was placed in two
setups as a singleton inside the \textit{H. neanderthalensis} skulls
(number $46$ in Fig. \ref{figNowsiolkaCloser_2}).
No particular pattern was observed in the setups related to the above,
in particular there seems to be no dependence on the distance function
($20$ are Sto{\l }yhwo's, $23$ $L^{2}$ and $22$ DD distances). 
However, some observations from the exemplary graphs in Figs. \ref{figNowsiolkaCloser_1} and \ref{figNowsiolkaCloser_2},
are presented in the Discussion. 
All the results are published alongside the source code. 
We also tested, what the results would be on the raw data, i.e. without normalization.
The Nowosi{\'o}{\l }ka skull would be placed more firmly within the \textit{H. sapiens}
skulls---in particular no singleton followed by \textit{H. neanderthalensis} skulls
was observed. 


\section{Discussion}
The work here has two aims. To show what possibilities \pkg{RMaCzek} has when creating
Czekanowski's diagrams and also to return to the original data that was used to introduce
the method to the world. Since Czekanowski used the method to challenge Sto{\l }yhwo's
conclusions concerning the evolutionary relatedness of the Nowosi{\'o}{\l }ka skull
we wanted to see which claim would modern methods substantiate and also
what approaches could support the alternative claim. 

Using various distance functions, variables and other settings
it was noticed that the placement of the Nowosi{\'o}{\l }ka
skull can greatly vary. However, in only $18$ out of the $96$ 
considered settings it can be thought to be classified as 
\textit{H. neanderthalensis} instead of \textit{H. sapiens}, in line 
with Sto{\l }yhwo's hypothesis. Interestingly none of these results, where
the Nowosi{\'o}{\l }ka skull is closest to the \textit{H. neanderthalensis},
were under Sto{\l }yhwo's distance function. 
Even in the cases (in Figs. \ref{figNowsiolkaCloser_1} and \ref{figNowsiolkaCloser_2}) where the Nowosi{\'o}{\l }ka skull was 
found closer to the \textit{H. neanderthalensis} skulls it was not grouped with them.
In setups $26$ and $92$ the Nowosi{\'o}{\l }ka skull is an outlier. 
In the other setups it is either just placed as bordering skull, displaying
weak similarity to the \textit{H. neanderthalensis} skulls (setup $82$),
or displaying similar similarity to \textit{H. sapiens} and \textit{H. neanderthalensis} skulls
(setups $13$, $21$, $66$). Alternatively, it was grouped with a \textit{H. sapiens} skull
in the midst of \textit{H. neanderthalensis} skulls (setup $29$).

From the multitude of obtained permutations, 
one must conclude that with so
few observations and high level of missing values the problem is very sensitive. 
It is of course impossible to completely replicate the analyses
from over $110$ years ago. In particular we do not know how Sto{\l }yhwo
classified each variable with respect to what is said about the Nowosi{\'o}{\l }ka skull's
placement.
Perhaps
with a different choice of skulls to be the ``standard''
we would have obtained a different result.

It must be pointed out that in this work we have restrained ourselves to 
data used by both Czekanowski and Sto{\l }yhwo. 
It is possible that they could have also tacitly supported their works with
data from other contemporary to them sources, e.g. \cite{GSch1906}.
Given today's technology one should compare the extractable
DNA from the Nowosi{\'o}{\l }ka skull with those of 
Neanderthals' and \textit{H. sapiens'}. This would definitely solve
the problem. We are not aware if this particular skull has
been sequenced, but certainly this would be beyond the scope
of this work, even if the actual specimen is still available somewhere.

While we do not have as our aim to draw conclusions concerning
archaic human populations it is worth commenting that our analyses do not contradict Czekanowski. 
The, Scythian, Nowosi{\'o}{\l }ka skull, seems to be related to the 
\textit{H. sapiens} skulls under most settings.
On the other hand, Sto{\l }yhwo's hypothesis was that neanderthalic
features did survive into the Era of History, perhaps strongly deformed.
What we know today, is that neanderthalic DNA is present in 
non--sub--Saharan modern human populations \cite{SSanetal2014},
proving, after a whole century, 
Sto{\l }yhwo correct, albeit in a way he could not had have foreseen at the time.

\section{Software Availability}
\pkg{RMaCzek} can be found at \url{https://cran.r-project.org/web/packages/RMaCzek/}
and \url{https://github.com/krzbar/RMaCzek/}. 
The \proglang{R} scripts allowing for the replication of the work here 
and the craniometric measurements from \cite{KSto1908}
are available at \url{https://github.com/krzbar/RMaCzek_KKZMBM2020}. 
The above scripts show how to
code user defined distance functions for further usage by \pkg{RMaCzek}.

\section{Acknowledgments} KB is grateful to Arkadiusz So{\l }tysiak for valuable comments.

\section{References}

\bibliography{\jobname}
\bibliographystyle{abbrvnat}

\bigskip
\setcounter{section}{0}
\selectlanguage{polish}
\Polskitrue
\subjclass[2010]{62H99; 62-04; 92B10} 
\keywords{diagram czekanowskiego, kraniometria, metody odleg{\l }o{\'s}ci wielocechowych, rozw{\'o}j ludzko{\'s}ci}
%

\begin{center}
{\bf Analiza czaszki z Nowosi{\'o}{\l }ki z u{\.z}yciem pakietu \pkg{RMaCzek}.}\\ 
\href{\repo/6476}{Krzysztof Bartoszek}
\end{center}
\medskip

\begin{abstract}
Na pocz{\k a}tku zesz{\l }ego stulecia Jan Czekanowski, polski antropolog oraz statystyk,
zaproponowa{\l\ } jedn{\k a} 
z pierwszych obiektywnych metod uporz{\k a}dkowania oraz zobrazowania macierzy odleg{\l }o{\'s}ci.
W $2019$ roku
zosta{\l } opracowany oraz udost{\k e}pniony pakiet \pkg{RMaCzek}, kt{\'o}ry pozwala na tworzenie
diagram{\'o}w Cze\-ka\-now\-skiego w {\'s}rodowisku \proglang{R}. W niniejszej pracy 
dokonano ponownej analizy danych, kt{\'o}re pos{\l }u{\.z}y{\l }y Czekanowskiemu do 
zaprezentowania w{\l }asnej metody oraz zaproponowano, jak w pakiecie \pkg{RMaCzek}
u{\.z}ytkownik mo{\.z}e wprowadza{\'c} w{\l }asn{\k a} funkcj{\k e} odleg{\l }o{\'s}ci.
\end{abstract}

\selectlanguage{polish}
\Polskitrue
\vspace{-3ex}
\begin{minipage}[b]{\linewidth}\small
\begin{wrapfigure}{l}{2.6cm}
  \vspace{-10pt}
  \begin{center}
    \includegraphics[width=2.6cm]{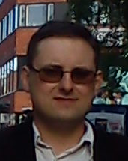}
  \end{center}
  \vspace{-10pt}
\end{wrapfigure}
\noindent \emph{Krzysztof Bartoszek}\footnote{\selectlanguage{english}References to his research papers are found in MathSciNet under  \href{http://www.ams.org/mathscinet/search/author.html?mrauthid= 793554}{ID: 793554} and the European Mathematical Society, FIZ Karlsruhe, and the Heidelberg Academy of Sciences bibliography database known as zbMath under \href{https://zbmath.org/authors/?q=ai: 	bartoszek.krzysztof}{ai:Bartoszek.Krzysztof}.} (ur. $1984$ w Bydgoszczy), obecnie wykładowca statystyki na Uniwersytecie w Link\"opingu. Absolwent informatyki Politechniki Gdańskiej (mgr inż.), biologii obliczeniowej Uniwersytetu w Cambridge (MPhil). Doktorat z statystyki matematycznej pod kierunkiem Serika Sagitova uzyskał w $2013$ r. na Uniwersytecie w G\"oteborgu. Jego główne zainteresowania związane są z procesami stochastycznymi w filogenetyce.
\hspace{2cm} 
\end{minipage}
\vspace{3ex}
\label{koniec}
{\Koniec}

\end{document}